\documentclass{ifacconf}

\usepackage{graphicx}      
\usepackage{amsmath,amssymb,bm}
\usepackage{natbib}        
\usepackage{tikz}
\usepackage{caption}
\usepackage{subfig}

\usepackage{algorithm}
\usepackage[noend]{algpseudocode}

\newtheorem{definition}{Definition}
\newtheorem{theorem}{Theorem}

\newtheorem{remark}{Remark}

\usepackage{./preamble/color-edits} 

\addauthor{ms}{purple}
\addauthor{ja}{blue}

\begin{document}
\begin{frontmatter}
\title{Auction-Based Responsibility Allocation for Scalable Decentralized Safety Filters in Cooperative Multi-Agent Collision Avoidance} 

\author[First]{Johannes Autenrieb} 
\author[First]{Mark Spiller} 

\address[First]{German Aerospace Center (DLR), Institute of Flight Systems, 38108, Braunschweig, Germany (email: \texttt{johannes.autenrieb@dlr.de, mark.spiller@dlr.de}).}

\begin{abstract}
This paper proposes a scalable decentralized safety filter for multi-agent systems based on high-order control barrier functions (HOCBFs) and auction-based responsibility allocation. While decentralized HOCBF formulations ensure pairwise safety under input bounds, they face feasibility and scalability challenges as the number of agents grows. Each agent must evaluate an increasing number of pairwise constraints, raising the risk of infeasibility and making it difficult to meet real-time requirements. To address this, we introduce an auction-based allocation scheme that distributes constraint enforcement asymmetrically among neighbors based on local control effort estimates. The resulting directed responsibility graph guarantees full safety coverage while reducing redundant constraints and per-agent computational load. Simulation results confirm safe and efficient coordination across a range of network sizes and interaction densities.
\end{abstract}


\end{frontmatter}

\section{Introduction}
Coordinated autonomy has become central to modern aerospace and robotic systems, enabling complex missions under tight spatial and temporal constraints. Applications such as distributed surveillance, formation flying, and cooperative engagement require multiple agents to operate in close proximity while maintaining safety and performance. These scenarios demand decentralized control architectures that ensure collision avoidance under bounded actuation and limited communication.

In cooperative missions,such as UAV swarms, interceptor formations, or autonomous vehicle fleets, simultaneous proximity maneuvers are unavoidable. While many existing assignment and guidance methods can coordinate task execution across agents~\citep{LI2024109212,Li2023}, they often overlook local interaction dynamics~\citep{Lv2024}. Consequently, even globally consistent task assignments may produce conflicting trajectories, underscoring the need for decentralized safety mechanisms that account for neighborhood-level interactions.

Control Barrier Functions (CBFs) offer a formal framework to guarantee safety by enforcing forward invariance of constraint sets~\citep{Ames_2017}. Their generalization to High-Order CBFs (HOCBFs) extends these guarantees to systems with higher relative degrees~\citep{Xiao_2022hocbf}, making them applicable to a wide range of multi-agent settings. 
Although CBF-based methods have shown success in decentralized coordination~\citep{BORRMANN2015UAV}, scalability and feasibility remain major challenges: as network size grows, each agent must evaluate an increasing number of pairwise constraints, raising computational load and the risk of infeasibility under control limitations~\citep{Aloor_2025,Isaly2024OnTheFeas}.

\begin{figure}[!t]
    \centering
    \includegraphics[width=0.85\columnwidth]{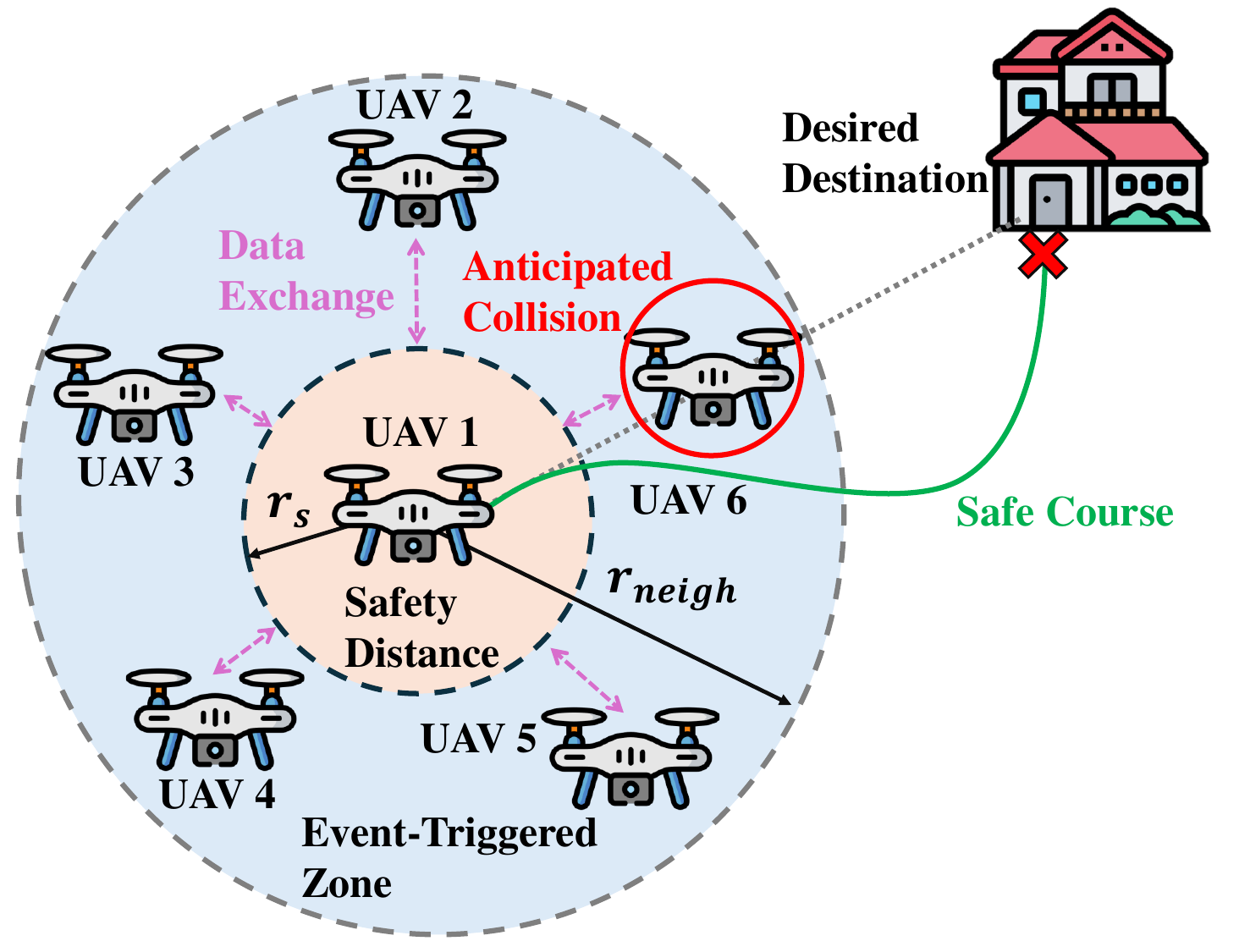}
    \caption{Illustration of the proposed neighborhood-based and auction-driven safety architecture.}
    \label{fig:overview_concept}
\end{figure}

To address these limitations, this paper introduces an auction-based responsibility allocation mechanism integrated into a decentralized safety filter architecture (see Fig.~\ref{fig:overview_concept}).  Agents coordinate which neighbor enforces each active constraint by exchanging lightweight cost bids representing their expected control deviation. The resulting responsibility graph assigns each active safety constraint to an enforcing agent with lower estimated control effort, thereby reducing redundant constraint evaluations and distributing safety responsibilities among agents.

The proposed architecture preserves the structure of the decentralized HOCBF formulation but operates on a reduced set of active constraints determined through distributed responsibility allocation. This improves real-time feasibility in dense environments, ensures complete safety coverage, and achieves scalable, coordinated behavior across tightly coupled multi-agent systems. Simulation results demonstrate that the method maintains safety as network size and interaction density increase.
\section{Preliminaries}
\label{sec:Preliminaries}
Consider a state space \( \chi \subset \mathbb{R}^n \) and a control input space \( U \subset \mathbb{R}^m \), where \( \chi \) is assumed to be path-connected and satisfies \( \mathbf{0} \in \chi \). Consider the control-affine nonlinear system
\begin{equation}
    \label{NonlinearPlant1}
    \dot{\mathbf{x}} = \mathbf{f}(\mathbf{x}) + \mathbf{g}(\mathbf{x}) \mathbf{u},
\end{equation}
where \( \mathbf{x} \in \chi \), \( \mathbf{u} \in U \), and \( \mathbf{f}: \chi \to \mathbb{R}^n \) and \( \mathbf{g}: \chi \to \mathbb{R}^{n \times m} \) are sufficiently smooth functions. To define safety, we consider a continuously differentiable function \( h: \chi \rightarrow \mathbb{R} \) and a set \( S \) defined as the zero-superlevel set of \( h \), yielding:
\begin{equation}
    \label{Safe_set_1}
    S \triangleq \big\{ \mathbf{x} \in \chi \mid h(\mathbf{x}) \geq 0 \big\}.
\end{equation}

\begin{definition}[Forward invariance and safety]
The set \( S \) is forward invariant for the system \eqref{NonlinearPlant1} if for every \( \mathbf{x}(0) \in S \), it follows that \( \mathbf{x}(t) \in S \) for all \( t \in I(\mathbf{x}_0) = [0, \tau_{\max} = \infty) \) \citep{Blanchini_1999}. The system \eqref{NonlinearPlant1} is safe with respect to \( S \) if \( S \) is forward invariant \citep{Ames_2017}.
\end{definition}

\begin{definition}[Class $\mathcal{K}$ functions, \citep{Khalil}]
A continuous function \( \alpha: (-b, a) \rightarrow \mathbb{R} \), with \( a,b > 0 \), is an extended class \( \mathcal{K} \) function (\( \alpha \in \mathcal{K} \)) if \( \alpha(0) = 0 \) and \( \alpha \) is strictly monotonically increasing. If \( a, b = \infty \) and \( \lim_{r \to \infty} \alpha(r) = \infty \), \( \lim_{r \to -\infty} \alpha(r) = -\infty \), then \( \alpha \) is a class \( \mathcal{K}_{\infty} \) function (\( \alpha \in \mathcal{K}_{\infty} \)).
\end{definition}

We introduce the notion of a \textit{control barrier function} (CBF) such that its existence allows the system to be rendered safe with respect to \( S \) using a control input \( \mathbf{u} \) \citep{Ames_2017}.
\begin{definition}[CBF, \citep{Ames_2017}]
Let \( S \subset \chi \) be the zero-superlevel set of a continuously differentiable function \( h: \chi \rightarrow \mathbb{R} \). The function \( h \) is a CBF for \( S \) for all \( \mathbf{x} \in S \), if there exists a class \( \mathcal{K}_{\infty} \) function \( \alpha(h(\mathbf{x})) \) such that for the dynamics defined in \eqref{NonlinearPlant1} we obtain:
\begin{equation}
    \label{ControlBarrierFunction_simple}
    \sup_{\mathbf{u}\in U} \dot{h}(\mathbf{x},\mathbf{u})  \geq -\alpha(h(\mathbf{x})),
\end{equation}
where
\begin{equation*}
\dot{h}(\mathbf{x},\mathbf{u}) = \frac{\partial h}{\partial \mathbf{x}}
\begin{bmatrix} 
\mathbf{f}(\mathbf{x}) + \mathbf{g}(\mathbf{x})\mathbf{u} 
\end{bmatrix}
= L_{\mathbf{f}} h(\mathbf{x}) + L_{\mathbf{g}} h(\mathbf{x}) \mathbf{u}.
\end{equation*}
\end{definition}

\begin{theorem}
\label{theorem_LCBF}
Given a set \( S \subset \chi \), defined via the associated CBF as in \eqref{Safe_set_1}, any Lipschitz continuous controller \( \mathbf{k}(\mathbf{x}) \in K_{S}(\mathbf{x}) \) with 
\begin{equation}
    K_{S} (\mathbf{x}) = \big\{ \mathbf{u} \in U : L_{\mathbf{f}} h(\mathbf{x}) + L_{\mathbf{g}} h(\mathbf{x}) \mathbf{u} + \alpha(h(\mathbf{x})) \ge 0 \big\}
    \label{definition_safe_controller}
\end{equation}
renders the system \eqref{NonlinearPlant1} forward invariant within \( S \) \citep{XU2015}.
\end{theorem}

For systems where the safety constraint \( h(\mathbf{x}) \) has a relative degree greater than one, the concept of control barrier functions can be generalized to high-order control barrier functions (HOCBFs) \citep{Xiao_2022hocbf}. 

\begin{definition}[HOCBF, \citep{Xiao_2022hocbf}]
Let \( h:\mathbb{R}^n \to \mathbb{R} \) be a \( d \)-times continuously differentiable function. Define recursively
\begin{align*}
    \psi_0(\mathbf{x}) &= h(\mathbf{x}),\notag\\
    \psi_1(\mathbf{x}) &= \dot{\psi}_0(\mathbf{x}) + \alpha_1(\psi_0(\mathbf{x})), \notag\\
    &\vdots \nonumber \notag\\
    \psi_d(\mathbf{x}) &= \dot{\psi}_{d-1}(\mathbf{x}) + \alpha_d(\psi_{d-1}(\mathbf{x})), \notag
\end{align*}

where $\alpha_1(\cdot), \alpha_2(\cdot), \ldots, \alpha_d(\cdot)$ are class \( \mathcal{K} \) functions. The function \( h(\mathbf{x}) \) is a \textit{High-Order Control Barrier Function} of relative degree \( d \) for the system \eqref{NonlinearPlant1} if there exist \( \alpha_1, \dots, \alpha_d \) such that
\begin{equation}
    L_{\mathbf{f}}^d h(\mathbf{x}) + L_{\mathbf{g}}^{d-1} h(\mathbf{x}) \mathbf{u} + O(h(\mathbf{x})) + \alpha_d(\psi_{d-1}(\mathbf{x})) \ge 0,
\end{equation}
\end{definition}
where \( O(h(\mathbf{x})) \) denotes the remaining Lie derivative terms along \( \mathbf{f} \), as well as possible partial derivatives with respect to time, of degree less than or equal to \( d-1 \).


\section{Problem Setup}
Let \( \mathcal{M} = \{1, 2, \dots, N\} \) denote the index set of \( N \) agents. The dynamics of each agent \( i \in \mathcal{M} \) are described by a second-order integrator model in three-dimensional Euclidean space. Specifically, the state vector \( x_i  \in \mathbb{R}^6 \) evolves according to the following linear time-invariant system:
\begin{equation}
    \dot{\mathbf x}_i = 
    \begin{bmatrix}
        \dot{\mathbf p}_i \\ \dot{\mathbf v}_i
    \end{bmatrix}
    =
    \begin{bmatrix}
        0 & \mathbf I \\
        0 & 0
    \end{bmatrix}
    \begin{bmatrix}
        \mathbf p_i \\ \mathbf v_i
    \end{bmatrix}
    +
    \begin{bmatrix}
        0 \\ \mathbf I
    \end{bmatrix}
    \mathbf a_i,
\label{eqn:double_integrator}
\end{equation}
where \( \mathbf{p}_i \in \mathbb{R}^3 \) denotes the position of agent \( i \), \( \mathbf v_i \in \mathbb{R}^3 \) its velocity, and \( \mathbf a_i \in \mathbb{R}^3 \) the control input representing the commanded acceleration vector.

\begin{remark}
The double-integrator model in \eqref{eqn:double_integrator} specifies the agent dynamics used in the subsequent safety-filter derivation. The auction-based allocation mechanism introduced later in this paper is not model-specific; it only requires that active pairwise safety conditions admit CBF or HOCBF representations in the admissible control variables. Thus, the same allocation perspective extends to other second-order, translational, control-affine models, including fixed-wing aircraft, once the corresponding safety-filter constraints and bid computation are derived
\end{remark}

The agents are assumed to be point masses subject to control and physical constraints, including input saturation.
We assume the following physical bounds:

\begin{equation}
    \|\mathbf v_i\| \leq \mathbf v_{\max}, \quad \|\mathbf a_i\| \leq \mathbf a_{\max}
\end{equation}

Define the relative quantities between two agents $i$ and $j$:
\begin{equation}
    \mathbf r_{ij} = \mathbf p_i - \mathbf p_j, \quad \mathbf v_{ij} = \mathbf v_i - \mathbf v_j
    \label{eq:realtive_kinematics}
\end{equation}

The mission objective is that each agent reaches an assigned static target destination, assumed to be externally provided and fixed throughout the mission. The guidance logic for each autonomy system is based on Proportional Navigation Guidance (PNG), which generates the nominal acceleration command $a_{\text{nom},i}$ required for successfully reaching an assigned target destination. 

Let the norm of the distance and the unit vector along the  line-of-sight (LOS) between the agent $i$ and the target $k$ be given by:
\begin{align}
    r &= \|\mathbf{r}_{ik}\| \\
    \hat{\mathbf{r}} &= \frac{\mathbf{r}_{ik}}{\|\mathbf{r}_{ik}\|}
\end{align}
The LOS angular rate vector is defined as:
\begin{equation}
    \dot{\boldsymbol{\lambda}} = \frac{\mathbf{r}_{ik} \times \mathbf{v}_{ik}}{\|\mathbf{r}_{ik}\|^2}.
\end{equation}

Each agent $i$ attempts to reach an assigned target $k$, using pure PNG. The pure PNG law prescribes a commanded acceleration perpendicular to the LOS direction, given by:
\begin{equation}
        \mathbf a_{\text{nom},i} = N \cdot \|\mathbf{v}_{ik}\| \cdot \left( \dot{\boldsymbol{\lambda}} \times \hat{\mathbf{r}} \right)
\end{equation}
where \( N > 0 \) is the navigation constant and \( \|\mathbf{v}_{ij}\| \) is the norm on relative closing speed. The acceleration vector lies orthogonal to the LOS and serves to nullify the LOS rate, aligning the velocity vector of the agent with the LOS over time.

While PNG provides efficient guidance commands, it does not account for interactions with neighboring agents. As a result, multiple agents pursuing different targets may inadvertently enter close proximity, increasing the risk of mutual collision. This motivates the integration of a real-time safety mechanism that modifies, for each agent \(i\), the nominal command \(\mathbf{a}_{\mathrm{nom},i}\) into a safe command \(\mathbf{a}_i\) while preserving feasibility and safety constraints.

\section{Neighborhood-Based Decentralized HOCBF Safety Filter Formulation}
As proposed in~\citep{BORRMANN2015UAV}, instead of considering all agents within the mission, each agent \( i \in \mathcal{M} = \{1,2,\dots,N\} \) restricts its safety evaluation to a local neighborhood to reduce computational complexity while maintaining safety guarantees. The neighborhood of agent \( i \) is defined as
\begin{equation}
\label{eq:neighborhood}
\mathcal{N}_i = \left\{ j \in \mathcal{M}\setminus\{i\} \; \middle| \; \|\mathbf{p}_i - \mathbf{p}_j\| \le r_{\text{neigh},i} \right\},
\end{equation}
where \( r_{\text{neigh},i} > 0 \) denotes the maximum interaction radius. The agent enforces safety constraints only with respect to neighbors within \( \mathcal{N}_i \), resulting in a decentralized formulation of the multi-agent safety problem.

To further reduce the burden of unnecessary constraint evaluation, an event-triggered scheme can be introduced as proposed in~\citep{autenrieb2025f}. In this framework, safety constraints with a neighbor \( j \in \mathcal{N}_i \) are activated only when two trigger conditions are simultaneously satisfied:

\begin{definition}[Event-triggered neighborhood]
The \emph{event-triggered neighborhood} (or \emph{active neighborhood}) of agent~$i$ is defined as the subset 
\(\mathcal{A}_i \subseteq \mathcal{N}_i\) containing all neighboring agents~$j$ for which a safety constraint is actively enforced.
An agent~$j$ belongs to the event-triggered neighborhood of~$i$, i.e. \(j \in \mathcal{A}_i\), if the following activation 
conditions are simultaneously satisfied:
\begin{itemize}
    \item \textbf{Proximity:} \( \|\mathbf{r}_{ij}\| \le r_{\text{crit},i} \), with \( 0 < r_{\text{crit},i} \le r_{\text{neigh},i} \);
    \item \textbf{Predicted zero-effort miss (ZEM):} \( \mathrm{ZEM}_{ij} \le \eta\, r_{\text{crit},i} \), with a design parameter \( \eta \in (0,1) \).
\end{itemize}
Here, \(r_{\text{crit},i}\) denotes the critical activation radius, \(r_{\text{neigh},i}\) the maximum interaction radius of the geometric neighborhood \(\mathcal{N}_i\), and \(\mathrm{ZEM}_{ij}\) the predicted minimum distance between agents~$i$ and~$j$ assuming constant relative velocity.
\label{def:active_neighborhood}
\end{definition}



The ZEM distance predicts the minimum separation between two agents assuming constant relative velocity. Considering $\mathbf{r}_{ij}$ and $\mathbf{v}_{ij}$, as defined in \eqref{eq:realtive_kinematics}, the relative position at time \( t=t_0+T \) evolves as
\begin{equation}
\mathbf{r}_{ij}(t) = \mathbf{r}_{ij}(t_0) + T\,\mathbf{v}_{ij}(t_0),
\end{equation}
and the time of closest approach is obtained by minimizing the squared distance \( D_{ij}^2(T)=\|\mathbf{r}_{ij}(t)\|^2 \), yielding
\begin{equation}
T_{\text{ZEM}} = -\frac{\mathbf{r}_{ij}(t_0)^\top \mathbf{v}_{ij}(t_0)}{\|\mathbf{v}_{ij}(t_0)\|^2}.
\end{equation}
If \( T_{\text{ZEM}} > 0 \), the agents are on a converging trajectory, and the predicted minimum separation is
\begin{equation}
\mathrm{ZEM}_{ij} = \|\mathbf{r}_{ij}(t_0) + T_{\text{ZEM}}\,\mathbf{v}_{ij}(t_0)\|.
\end{equation}

Whenever the activation condition is satisfied, a HOCBF is constructed from the perspective of agent \( i \) to ensure pairwise safety with respect to neighbor \( j \). For a minimum required separation \( r_s > 0 \), the barrier function for the zero-superlevel set \( \mathcal{S}_{ij}=\{x_i,x_j\in  \mathbb{R}^6 \mid h_{ij}(x_i,x_j)=\|\mathbf{r}_{ij}\|^2 - r_s^2\ge0\}\) is defined as
\begin{equation}
h_{ij}(\mathbf{x}_i, \mathbf{x}_j) = \|\mathbf{r}_{ij}\|^2 - r_s^2.
\end{equation}

The corresponding time derivatives follow from the relative dynamics
\begin{align}
\dot{h}_{ij} &= 2\,\mathbf{r}_{ij}^\top \mathbf{v}_{ij}, \\
\ddot{h}_{ij} &= 2\|\mathbf{v}_{ij}\|^2 + 2\,\mathbf{r}_{ij}^\top(\mathbf{a}_i - \mathbf{a}_j),
\end{align}
where \( \mathbf{a}_i \) denotes the acceleration command of agent \( i \), and \( \mathbf{a}_j \) represents the (unknown) control input of neighbor \( j \).

Following the formulation introduced in Sect.~\ref{sec:Preliminaries}, the recursive HOCBFs are given by
\begin{equation}
\psi_0 = h_{ij}, \quad
\psi_1 = \dot{\psi}_0 + \gamma_1 \psi_0, \quad
\psi_2 = \dot{\psi}_1 + \gamma_2 \psi_1,
\end{equation}
with class-\(\mathcal{K}\) gains \( \gamma_1, \gamma_2 > 0 \). The safety condition requires
\begin{equation}
\psi_2(\mathbf{x}_i, \mathbf{x}_j , \mathbf{a}_i, \mathbf{a}_j) \ge 0,
\end{equation}
which leads to the affine-in-control constraint
\begin{equation}
(2\,\mathbf{r}_{ij})^\top \mathbf{a}_i - (2\,\mathbf{r}_{ij})^\top \mathbf{a}_j \ge c_{ij}(\mathbf{x}_i, \mathbf{x}_j, \gamma_1, \gamma_2),
\end{equation}
where
\begin{equation}
\begin{split}
c_{ij}(\mathbf{x}_i, \mathbf{x}_j, \gamma_1, \gamma_2) =
&-2\|\mathbf{v}_{ij}\|^2
-2(\gamma_1+\gamma_2)\mathbf{r}_{ij}^\top \mathbf{v}_{ij}\\
&-\gamma_2(\|\mathbf{r}_{ij}\|^2 - r_s^2).
\end{split}
\end{equation}

Since agent \( i \) cannot access \( \mathbf{a}_j \) directly, assumptions on the neighbor’s behavior are required. When communication is available, a cooperative avoidance behavior is assumed with \( \mathbf{a}_j = -\mathbf{a}_i \), corresponding to both agents accelerating away from each other. In the absence of a communication link, a non-adversarial assumption is used with \( \mathbf{a}_j = \mathbf{0} \), representing a neighbor maintaining its trajectory without active avoidance. These assumptions allow each agent to independently evaluate its safety constraint without centralized coordination.

\begin{figure}[!t]
    \centering
    \includegraphics[width=0.9\columnwidth]{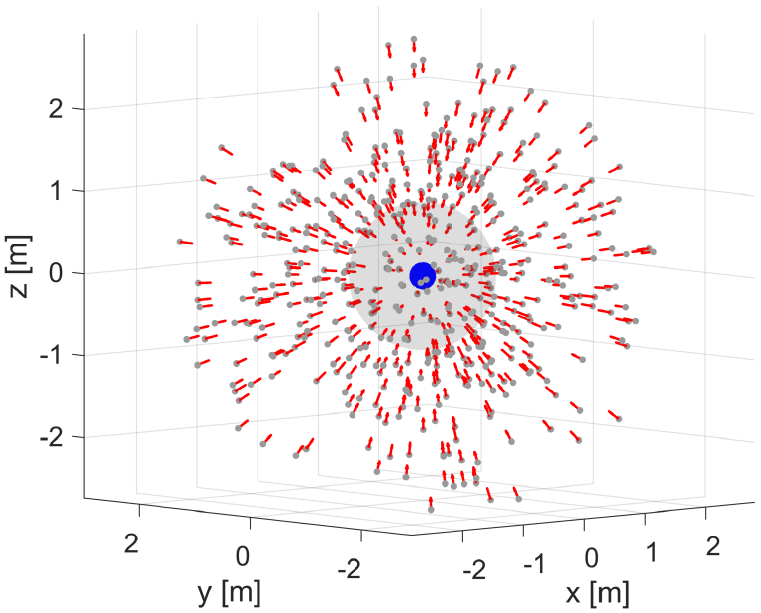}
    \caption{Evaluated pairwise collision scenarios for HOCBF parameter tuning.}
    \label{fig:gamma_feasibility}
\end{figure}

The choice for value of the HOCBF parameters $\gamma_1$ and $\gamma_2$ critically influence the admissible control authority required to maintain safety and thus determine feasibility under input constraints. Improper tuning may render the constraints overly conservative or infeasible, particularly when multiple agents operate in close proximity \citep{Xiao2020_feasibility}. To systematically characterize this dependence, a feasibility mapping was conducted through Monte Carlo simulations of pairwise collision scenarios. As illustrated in Fig.~\ref{fig:gamma_feasibility}, agent~$i$ (blue) was fixed at the origin and surrounded by a spherical safety boundary (gray), while agent~$j$ was randomly positioned along this boundary with velocity vectors directed toward~$i$ (red arrows). Each sampled configuration represents a distinct encounter geometry for which the safety filter was solved to determine whether a feasible control input satisfying the HOCBF constraint existed. This evaluation provides a data-driven guideline for selecting $(\gamma_1,\gamma_2)$ values that ensures feasibility across diverse interaction cases while maintaining sufficient responsiveness for collision avoidance. The identified parameter range is subsequently applied in the decentralized multi-agent simulations to provide consistent safety performance under input constraints.

\begin{remark}
The analysis for evaluating and tuning the integrated HOCBF formulation is conducted under a one-on-one collision avoidance scenario. It should be emphasized that this represents only a sub-approximation of the full multi-agent problem, where an HOCBF would, in principle, need to be derived for every possible interaction scenario. Nevertheless, it can be reasonably assumed that, in most cases, the individual pairwise safety constraints derived from the one-on-one setting can be dynamically integrated as separate constraints within the decentralized safety filter. Moreover, we will later show that a responsibility allocation scheme can help to minimize the risk of potentially conflicting independent safety constraints.
\end{remark}

The safe control command for each agent \( i \) is obtained by solving a local quadratic program that enforces all active pairwise constraints with neighbors \( j \in \mathcal{A}_i \subseteq \mathcal{N}_i \):

\begin{align}
\min_{\mathbf{a}_i \in \mathbb{R}^3} \quad
& \|\mathbf{a}_i - \mathbf{a}_{\text{nom},i}\|^2 \\
\text{s.t.} \quad
& (2\,\mathbf{r}_{ij})^\top \mathbf{a}_i - (2\,\mathbf{r}_{ij})^\top \mathbf{a}_j \ge c_{ij}(\mathbf{x}, \gamma_1, \gamma_2), \quad \forall j \in \mathcal{A}_i, \notag\\
& \mathbf{a}_{\min} \le \mathbf{a}_i \le \mathbf{a}_{\max}.\notag
\end{align}
This decentralized formulation allows each agent to evaluate and enforce its own safety constraints using only locally available information. The assumptions on \( \mathbf{a}_j \) ensure that safety can be maintained under both cooperative and communication-limited conditions.




However, as the number of agents increases, the number of active pairwise constraints per agent may grow, which can increase computational complexity. In dense environments, this can potentially lead to a loss of real-time guarantees. Moreover, with a growing number of agents, the likelihood of conflicting constraints increases, which can result in feasibility issues. In the following section, an auction-based decentralized method is introduced to address these challenges by reducing the number of active constraints for each agent, while maintaining overall safety guarantees.
\section{Auction-Based Responsibility Allocation for Scalable Decentralized Safety Filter}
In the previous section, the decentralized event-triggered safety filter allowed each agent to guarantee pairwise safety within its local neighborhood. However, as the number of agents increases, the number of simultaneously active constraints per agent grows, which can lead to redundant constraint evaluations, increased computational burden, and potential feasibility loss. 

To address these challenges, we introduce an additional coordination layer on top of the decentralized safety filter. The goal is to distribute the responsibility for constraint enforcement across the network such that all relevant safety conditions remain satisfied, but each agent only enforces a subset of them. This reduces redundant computations and ensures that every safety-relevant interaction is covered by at least one enforcing agent. Beyond computational advantages, this approach also aligns the control objective with the overall mission-level goal: rather than minimizing the deviation from the nominal command individually for each agent, the system aims to minimize the cumulative deviation across all agents, thereby maintaining collective optimality and consistent global behavior. From a control-theoretic standpoint, this corresponds to an optimal responsibility allocation problem over a directed interaction graph, where the directed edge \( (i \rightarrow j) \) denotes that agent \( i \) is responsible for ensuring safety with respect to agent \( j \).

We retain the geometric neighborhood definition~\(\mathcal{N}_i\) and the event-triggered activation logic introduced in the previous section (see Definition~\ref{def:active_neighborhood}). 
Let \(\mathcal{A}_i \subseteq \mathcal{N}_i\) denote the event-triggered neighborhood of agent~\(i\), representing all currently active neighbors for which safety constraints are enforced. 
Each agent may take responsibility for enforcing safety only with respect to a subset \(\mathcal{S}_i \subseteq \mathcal{A}_i\), which defines a directed responsibility graph
\begin{equation}
\mathcal{G}_S = (\mathcal{M}, \bigcup_{i \in \mathcal{M}} \mathcal{E}_{S_i}),
\qquad 
\mathcal{E}_{S_i} = \{ (i \rightarrow j) \mid j \in \mathcal{S}_i \}.
\end{equation}
The coverage condition requires that, for every active pair \((i,j)\), at least one of the two agents is responsible for enforcing the corresponding safety constraint, i.e.
\begin{equation}
\forall (i,j) \in \mathcal{A}_i \cup \mathcal{A}_j : 
\ (i \rightarrow j) \in \mathcal{E}_{S_i} \ \text{or} \ (j \rightarrow i) \in \mathcal{E}_{S_j}.
\label{eq:coverage_cond}
\end{equation}
This ensures that all relevant safety interactions are covered by at least one enforcing agent, without requiring symmetric enforcement of constraints.

A centralized abstraction of this responsibility allocation can be expressed as the following combinatorial optimization problem. 
Let \(\bm z = \{z_{ij} \mid i \in \mathcal{M},\, j \in \mathcal{A}_i\}\) denote the set of binary assignment variables, 
where \(z_{ij}=1\) indicates that agent~\(i\) is responsible for enforcing the safety constraint with respect to neighbor~\(j\), 
and \(z_{ij}=0\) otherwise. The parameter \(\mathcal{C}_i\) bounds the number of simultaneously active responsibilities of agent~\(i\), 
and \(C_{ij}\) quantifies the enforcement cost associated with agent~\(i\) being responsible for agent~\(j\). 
The centralized formulation then reads
\begin{align}
\min_{\bm z}&\;\sum_{i=1}^N\sum_{j\in\mathcal{A}_i} C_{ij}\,z_{ij}
\quad\notag\\
\text{s.t.} \quad
& z_{ij}\in\{0,1\}, \label{eq:central_assign_dir}\\
& \sum_{j} z_{ij} \le \mathcal{C}_i, \notag\\
& (i,j) \in \mathcal{A}_i \cup \mathcal{A}_j.\notag
\end{align}

Unlike a symmetric assignment where \(z_{ij}=z_{ji}=1\), the directed responsibility formulation in~\eqref{eq:coverage_cond} 
allows asymmetric allocations such as \(1\!\rightarrow\!2\), \(2\!\rightarrow\!3\), \(3\!\rightarrow\!1\), 
which still ensure complete safety coverage while minimizing redundancy. 
Solving~\eqref{eq:central_assign_dir} centrally would yield a globally optimal configuration 
but is incompatible with the decentralized communication structure. 
To achieve a comparable effect locally, we approximate~\eqref{eq:central_assign_dir} 
through a distributed, greedy, auction-based consensus mechanism 
that converges to a consistent responsibility graph~\(\mathcal{G}_S\) 
using only local communication among neighboring agents~\citep{BRAQUET2021}.

To define a computable and comparable local cost, we recall the HOCBF condition from Sect.~\ref{sec:Preliminaries},
\begin{equation}
(2\,\bm r_{ij})^\top \bm a_i - (2\,\bm r_{ij})^\top \bm a_j \;\ge\; c_{ij}(\bm x,\gamma_1,\gamma_2), \label{eq:cbf_affine_keep}
\end{equation}
and replace the unknown \(\bm a_j\) by its admissible estimate \(\bar{\bm a}_j\) to obtain
\begin{equation}
\bm (2\,\bm r_{ij})^\top \bm a_i \;\ge\; b_{ij}, \qquad
b_{ij}:=c_{ij} + (2\,\bm r_{ij})^\top \bar{\bm a}_j . \label{eq:halfspace_keep}
\end{equation}
For this single active constraint, the one-constraint safety filter reads
\begin{equation}
\min_{\bm a_i}\ \|\bm a_i-\bm a_{\text{nom},i}\|^2
\quad \text{s.t.}\quad \bm (2\,\bm r_{ij})^\top \bm a_i \ge b_{ij}. \label{eq:proj_qp}
\end{equation}
Using the KKT conditions, we obtain the closed-form projection
\begin{equation}
\bm a_i^\star \;=\; \bm a_{\text{nom},i} +
\frac{(b_{ij}-\bm (2\,\bm r_{ij})^\top \bm a_{\text{nom},i})_+}{\|\bm 2\,\bm r_{ij}\|^2}\,\bm (2\,\bm r_{ij}), \label{eq:proj_closed}
\end{equation}
with $(x)_+:=\max\{x,0\}$ and the associated deviation cost
\begin{equation}
\|\bm a_i^\star-\bm a_{\text{nom},i}\|^2 \;=\;
\frac{\bigl(b_{ij}-\bm (2\,\bm r_{ij})^\top \bm a_{\text{nom},i}\bigr)_+^2}{\|\bm 2\,\bm r_{ij}\|^2}. \label{eq:dev_cost}
\end{equation}

This term quantifies the minimal corrective effort required for agent \(i\) to render constraint \((i,j)\) safe. Under homogeneous agent assumptions (i.e., comparable nominal commands and authority limits), the term can serve as a fast and consistent proxy for the true deviation cost \(C_{ij}\) in \eqref{eq:central_assign_dir}, thereby enabling the evaluation of the relative enforcement effort across agents. We therefore define the local bid function
\begin{equation}
J_{ij}^{(i)} \;:=\; \frac{\bigl(b_{ij}-\bm (2\,\bm r_{ij})^\top \bm a_{\text{nom},i}\bigr)_+^2}{\|\bm 2\,\bm r_{ij}\|^2}, 
\label{eq:bid_keep}
\end{equation}
with
\begin{equation*}
\qquad
J_{ij}^{(i)}:=+\infty\ \text{if }\|\bm a_i^\star\|_\infty>a_{\max}, 
\end{equation*}
and where the superscript \((i)\) indicates that the corrective effort is evaluated from the perspective of agent \(i\). The bid \(J_{ij}^{(i)}\) corresponds exactly to the optimal value of the projection problem~\eqref{eq:proj_qp}, providing a lower bound on the deviation cost that would arise if the constraint were included in the full safety filter.

All agents within the active neighborhood \(\mathcal{V}_i := \mathcal{A}_i \cup \{i\}\) exchange bids for all active links \((p,q)\in \mathcal{V}_i\) and execute a local auction to determine the direction of responsibility assignment:
\begin{equation}
\sigma_{pq} \;=\; \arg\min_{k \in \{p,q\}} J_{pq}^{(k)}
\quad \text{s.t.}\quad |\mathcal{S}_k|\le C_k,
\label{eq:winner_rule}
\end{equation}
where \(\mathcal{S}_k\) denotes the set of neighbors for which agent \(k\) is responsible. The result of the distributed auction process defines the directed responsibility edges \((k \to j)\) in \(\mathcal{G}_S\), satisfying the coverage condition \eqref{eq:coverage_cond}. If all bids are infeasible, temporary dual enforcement is applied until feasibility is restored.


Let \(\mathcal{F}_i\subseteq\mathcal{M}\) denote forced-zone neighbors that must always be included for safety-critical proximity. The final active constraint set for agent \(i\) is then given by \(\mathcal{A}^\Sigma_i=\mathcal{S}_i\cup\mathcal{F}_i\). The safe acceleration command is obtained using the same quadratic program as in the previous section, now with the reduced constraint set:
\begin{align}
\min_{\bm a_i\in\mathbb{R}^3}\quad & \|\bm a_i-\bm a_{\text{nom},i}\|^2 \label{eq:qp_same_form}\\
\text{s.t.}\quad
& (2\,\bm r_{ij})^\top \bm a_i - (2\,\bm r_{ij})^\top \bm a_j \;\ge\; c_{ij}(\bm x,\gamma_1,\gamma_2),\hspace{0.25em}\forall j\in\mathcal{A}^\Sigma_i,\notag\\
& \bm a_{\min}\le \bm a_i \le \bm a_{\max}.\notag
\end{align}
This ensures that every active safety constraint is covered by at least one enforcing agent, while the number of simultaneously active constraints per agent is minimized. 

\section{Results}
\begin{figure*}[!t]
    \centering
    \subfloat[Three agents]{%
        \includegraphics[width=0.32\textwidth]{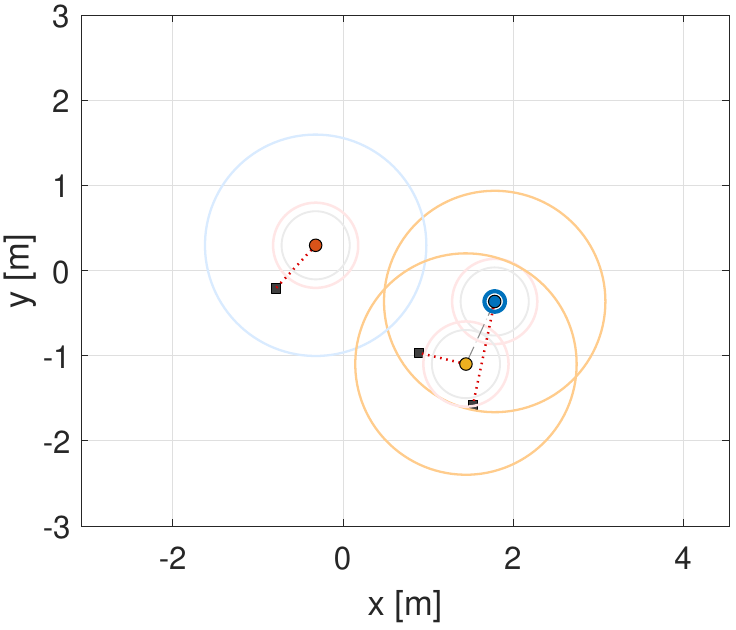}%
        \label{fig:sim3agents}
    }\hfill
    \subfloat[Eight agents]{%
        \includegraphics[width=0.32\textwidth]{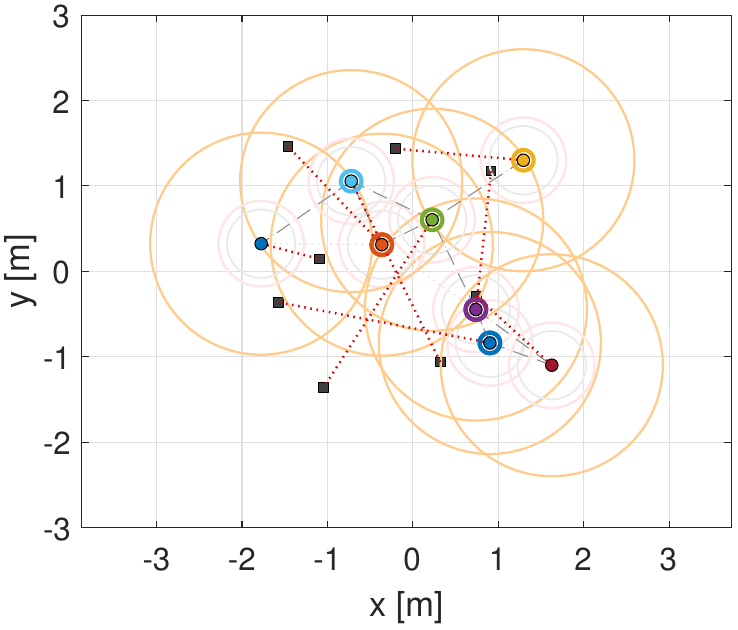}%
        \label{fig:sim8agents}
    }\hfill
    \subfloat[Twenty agents]{%
        \includegraphics[width=0.32\textwidth]{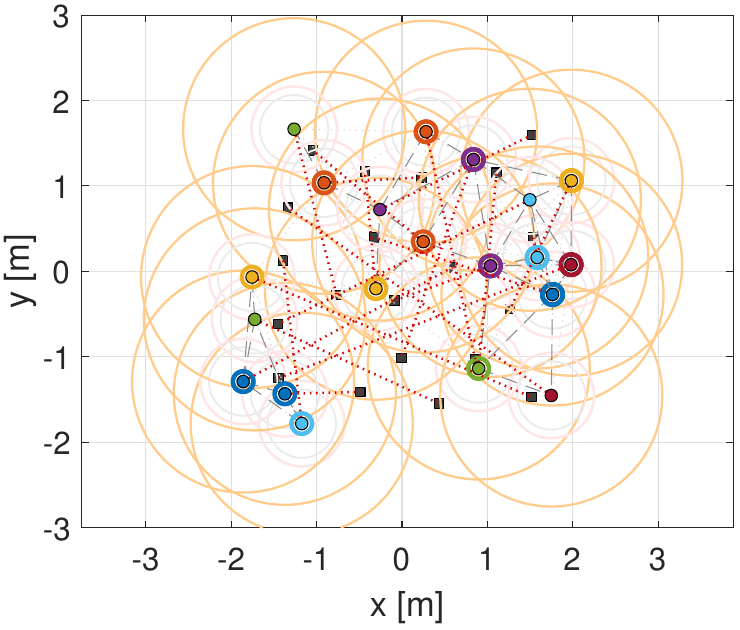}%
        \label{fig:sim20agents}
    }
    \caption{Comparison of decentralized HOCBF-based safety performance for different numbers of agents.}
    \label{fig:sim_overview}
\end{figure*}

A MATLAB-based simulation environment was developed to evaluate the proposed responsibility-allocation-driven decentralized HOCBF-based safety filter under various multi-agent conditions. All agents were modeled as double integrators following \eqref{eqn:double_integrator}, with the dynamics restricted to the two-dimensional plane for this scenario. Each agent was assigned an independent target position, indicated by a black square. The red dash-dotted lines indicate the LOS between the agents and their assigned targets. Each agent maintains a circular safety zone of radius \(r_s = 0.4~\mathrm{m}\) and an event-triggered neighborhood radius of \(r_{\mathrm{neigh}} = 1.3~\mathrm{m}\), while inter-agent communication is permitted within \(r_{\mathrm{comm}} = 1.6~\mathrm{m}\). The agents operate in a bounded \(6\times6~\mathrm{m}\) workspace, forming a dense and dynamically coupled interaction environment designed to challenge both the scalability and feasibility of the decentralized safety architecture.

Figure~\ref{fig:sim_overview} compares three representative cases with different network densities: (a) three agents, (b) eight agents, and (c) twenty agents. The gray dashed lines denote active safety constraints between agents, while thick circles around an agent indicate active responsibility. Blue and orange safety-zone circles indicate inactive and active collision avoidance evaluation routines, respectively. As shown in Fig.~\ref{fig:sim3agents}, when two agents approach each other, the decentralized safety mechanism is activated to maintain the required minimum separation, while isolated agents with no imminent conflicts continue toward their targets unaltered. With an increasing number of agents, as in Figs.~\ref{fig:sim8agents} and~\ref{fig:sim20agents}, the network exhibits more complex but coordinated activation patterns, demonstrating that the proposed approach maintains safe multi-agent operation under dense interaction conditions.

\section{Conclusion}
This paper presented a decentralized safety-critical control framework based on high-order control barrier functions combined with an auction-based responsibility allocation mechanism. By distributing safety responsibilities through local consensus, each agent enforces only a reduced set of constraints while full network coverage and formal safety guarantees are preserved. The approach minimizes redundant constraint evaluations, improves feasibility under dense interactions, and ensures near-optimal collective performance through cost-based responsibility assignment. Numerical simulation results for different numbers of agents demonstrate that the proposed method maintains safe inter-agent interaction.

\bibliography{ifacconf}

@Article{Lv2024,
AUTHOR = {Lv, Xinyuan and Peng, Chi and Ma, Jianjun},
TITLE = {Control Barrier Function-Based Collision Avoidance Guidance Strategy for Multi-Fixed-Wing UAV Pursuit-Evasion Environment},
JOURNAL = {Drones},
VOLUME = {8},
YEAR = {2024},
NUMBER = {8},
ARTICLE-NUMBER = {415},
DOI = {10.3390/drones8080415}
}

@article{LI2024109212,
title = {A comprehensive survey of weapon target assignment problem: Model, algorithm, and application},
journal = {Engineering Applications of Artificial Intelligence},
volume = {137},
pages = {109212},
year = {2024},
issn = {0952-1976},
doi = {10.1016/j.engappai.2024.109212}    ,
author = {Jinrui Li and Guohua Wu and Ling Wang}
}

@INPROCEEDINGS{Li2023,
  author={Li, Xuheng and Yu, Jianglong and Dong, Xiwang and Ren, Zhang and Feng, Zhuo and Liang, Biao},
  booktitle={2023 42nd Chinese Control Conference (CCC)}, 
  title={Impact Time Cooperative Guidance for Multi-Missile System Based on Incremental Learning}, 
  year={2023},
  volume={},
  number={},
  pages={3766-3771},
  doi={10.23919/CCC58697.2023.10239914}}

@ARTICLE{Xiao_2022hocbf,
  author={Xiao, Wei and Belta, Calin},
  journal={IEEE Transactions on Automatic Control}, 
  title={High-Order Control Barrier Functions}, 
  year={2022},
  volume={67},
  number={7},
  pages={3655-3662},
  doi={10.1109/TAC.2021.3105491}}

@inproceedings{Aloor_2025,
author = {Aloor, Jasmine and Choi, Jason and Li, Jingqi and Mendoza, Maria and Balakrishnan, Hamsa and Tomlin, Claire},
year = {2025},
month = {06},
pages = {},
booktitle = {Robotics: Science and Systems Conference},
title = {Resolving Conflicting Constraints in Multi-Agent Reinforcement Learning with Layered Safety},
doi = {10.15607/RSS.2025.XXI.094}
}

@ARTICLE{Ames_2017,
  author={Ames, Aaron D. and Xu, Xiangru and Grizzle, Jessy W. and Tabuada, Paulo},
  journal={IEEE Transactions on Automatic Control}, 
  title={Control Barrier Function Based Quadratic Programs for Safety Critical Systems}, 
  year={2017},
  volume={62},
  number={8},
  pages={3861-3876},
  doi={10.1109/TAC.2016.2638961  }}

@article{XU2015,
title = {Robustness of Control Barrier Functions for Safety Critical Control},
journal = {IFAC-PapersOnLine},
volume = {48},
number = {27},
pages = {54-61},
year = {2015},
issn = {2405-8963},
doi = {10.1016/j.ifacol.2015.11.152},
author = {Xiangru Xu and Paulo Tabuada and Jessy W. Grizzle and Aaron D. Ames}
}

@book{Khalil,
author        = "Khalil, Hassan K",
title         = "{Nonlinear systems; 3rd ed.}",
publisher     = "Prentice-Hall",
address       = "Upper Saddle River, NJ",
year          = "2002",
}

@article{Blanchini_1999,
title = {Set invariance in control},
journal = {Automatica},
volume = {35},
number = {11},
pages = {1747-1767},
year = {1999},
issn = {0005-1098},
doi = {10.1016/S0005-1098(99)00113-2},
author = {F. Blanchini}
}

@ARTICLE{Isaly2024OnTheFeas,
	author={Isaly, Axton and Ghanbarpour, Masoumeh and Sanfelice, Ricardo G. and Dixon, Warren E.},
	journal={IEEE Transactions on Automatic Control}, 
	title={On the Feasibility and Continuity of Feedback Controllers Defined by Multiple Control Barrier Functions}, 
	year={2024},
	volume={},
	number={},
	pages={1-15},
	doi={10.1109/TAC.2024.3383069}}

@article{BORRMANN2015UAV,
title = {Control Barrier Certificates for Safe Swarm Behavior},
journal = {IFAC-PapersOnLine},
volume = {48},
number = {27},
pages = {68-73},
year = {2015},
issn = {2405-8963},
doi = {10.1016/j.ifacol.2015.11.154} ,
author = {Urs Borrmann and Li Wang and Aaron D. Ames and Magnus Egerstedt}
}

@INPROCEEDINGS{Xiao2020_feasibility,
  author={Xiao, Wei and Belta, Calin A. and Cassandras, Christos G.},
  booktitle={2020 59th IEEE Conference on Decision and Control (CDC)}, 
  title={Feasibility-Guided Learning for Constrained Optimal Control Problems}, 
  year={2020},
  volume={},
  number={},
  pages={1896-1901},
  doi={10.1109/CDC42340.2020.9303857}}

@article{BRAQUET2021,
title = {Greedy Decentralized Auction-based Task Allocation for Multi-Agent Systems},
journal = {IFAC-PapersOnLine},
volume = {54},
number = {20},
pages = {675-680},
year = {2021},
note = {2021 Modeling, Estimation and Control Conference (MECC)},
issn = {2405-8963},
doi = {10.1016/j.ifacol.2021.11.249}  ,
author = {Martin Braquet and Efstathios Bakolas}
}

@misc{autenrieb2025f,
      title={Decentralized CBF-based Safety Filters for Collision Avoidance of Cooperative Missile Systems with Input Constraints}, 
      author={Johannes Autenrieb and Mark Spiller},
      year={2025},
      eprint={2510.06846},
      archivePrefix={arXiv},
      primaryClass={eess.SY},
      url={https://arxiv.org/abs/2510.06846} 
}

\end{document}